\documentclass[useAMS,usenatbib]{mnras}

\usepackage{amsmath,amsfonts,graphicx,txfonts,fleqn,color,tabularx,verbatim,fixltx2e}
\newcommand {\B}[1] {\boldsymbol{#1}}

\newcommand  *{\D}[1] {\dot{\B{#1}}}
\newcommand  *{\DD}[1] {\ddot{\B{#1}}}

\title[Breaking mean-motion resonances]{Breaking mean-motion resonances during Type I planet migration}
\author[Hands \& Alexander]{T.O.Hands$^{1,2}$\thanks{email:
tomhands@physik.uzh.ch} and R.D.Alexander$^{1}$ 
\\$^{1}$Department of Physics \& Astronomy, University of Leicester, University Road, Leicester, LE1 7RH, UK
\\$^{2}$Institut f\"ur Computergest\"utzte Wissenschaften, Universit\"at Z\"urich, Winterthurerstrasse 190, 8057 Z\"urich, Switzerland}
\begin{document}
\voffset=-0.25in

\pagerange{\pageref{firstpage}--\pageref{lastpage}} \pubyear{2017}

\date{Accepted 2017 October 16; Submitted 2017 October 13; in original form 2017 March 27}

\maketitle

\label{firstpage}

\begin{abstract}
We present two-dimensional hydrodynamical simulations of pairs of planets migrating simultaneously in the Type I regime in a protoplanetary disc. Convergent migration naturally leads to the trapping of these planets in mean-motion resonances. Once in resonance the planets' eccentricity grows rapidly, and disc-planet torques cause the planets to escape resonance on a time-scale of a few hundred orbits. The effect is more pronounced in highly viscous discs, but operates efficiently even in inviscid discs. We attribute this resonance-breaking to overstable librations driven by moderate eccentricity damping, but find that this mechanism operates differently in hydrodynamic simulations than in previous analytic calculations. Planets escaping resonance in this manner can potentially explain the observed paucity of resonances in \textit{Kepler} multi-transiting systems, and we suggest that simultaneous disc-driven migration remains the most plausible means of assembling tightly-packed planetary systems.
\end{abstract}



\section{Introduction}
Multi-transiting exoplanetary systems provide a wealth of insights into the early evolution of young planetary systems. The architectures of these systems, which are typically observed at $\sim$Gyr ages, depend critically on the formation and evolution of planets in their parent protoplanetary discs, which typically live for $\lesssim 10$Myr \citep[e.g.,][]{Haisch2001,Fedele2010}. Subsequent secular evolution can of course modify these architectures, but how planetary systems evolve during the disc's lifetime remains a critical uncertainty in our understanding of planet formation.

Of particular interest are the class of compact, close-packed planetary systems discovered by the {\it Kepler} mission. These systems, for which Kepler-11 is the prototype \citep{Lissauer2011}, typically consist of four or more super-Earth- or Neptune-sized planets in very short period orbits, within a few tenths of an AU of their host stars \citep[e.g.,][]{Swift2013,Lissauer2013}. These planets typically orbit within a few hundredths of an AU of one another, and remain dynamically cold (with near-zero orbital inclinations and eccentricities) despite apparently being close to dynamical instability \citep{Deck2012}. The prevalence of these systems suggests a robust formation mechanism, yet they apparently require delicate assembly to avoid catastrophic dynamical instabilities.

Traditional core accretion theory holds that planets form at large distances from their host stars \citep[$\gtrsim$1--10AU, see e.g.,][]{Pollack1996,Raymond2014a} and migrate to smaller orbital radii through gravitational interactions with the protoplanetary disc \citep[e.g.,][]{Goldreich1980,Lin1986}. Planets in compact systems are typically low mass, and are therefore expected to migrate in the Type I regime \citep[see, e.g., the reviews by][and references therein]{Kley2012,Baruteau2014}. We have previously demonstrated that formation at AU radii followed by Type I migration is a viable mechanism for forming compact planetary systems such as Kepler-11 \citep{Hands2014,Hands2016}. However, simultaneous Type I migration of multiple planets invariably leads to an abundance of planets becoming trapped in mean-motion resonances \citep[MMRs; e.g.,][]{Cresswell2006}, which contrasts sharply with the lack of resonances observed in {\it Kepler} multiple systems \citep{Fabrycky2012}. To explain the observed systems we must therefore either identify a robust mechanism for breaking MMRs during Type I migration, or appeal to more extreme models such as {\it in situ} formation \citep[e.g.,][]{Hansen2012,Chiang2013}. Moreover, there is no reason that these models should be mutually exclusive, and it seems likely that even planets formed at sub-AU radii will migrate significantly during the Myr lifetime of the protoplanetary disc. Understanding how pairs of migrating planets can escape from MMRs is therefore a key ingredient in any model for the formation of compact planetary systems. 

Two planets $i$ and $j$ are said to be in the  $p : p + q$ mean-motion resonance if the resonant argument, given by
\begin{equation}
\psi = (p+q) \lambda_j - p \lambda_i - q \varpi_j,
\label{eq:resarg}
\end{equation}
librates, rather than circulates. Here, $\lambda = M + \varpi$ is the mean longitude, $M$ is the mean anomaly, $\varpi$ is the longitude of pericentre and $p$ and $q$ are integers \citep{MurrayDermott}. Note that one can replace $\varpi_j$ with $\varpi_i$ in the above equation, and instead look for libration in this second resonant angle. The two resonant angles may librate around the same or different mean values \cite[see e.g.][]{Lee2004}. Once in resonance the planet-planet torque typically dominates, and pairs (or chains) of resonant planets tend to migrate together while remaining trapped in resonance. Several mechanisms for breaking MMRs have been proposed, such as disc turbulence \cite[e.g.,][]{Terquem2007,Rein2012,Paardekooper2013}, planet-wake interactions \citep{Baruteau2013}, gravitational perturbations by unseen outer planets \citep{Hands2016}, one-sided migration torques during disk dispersal \citep{Liu2017}, tidal interactions with the star \citep{Papaloizou2011}, and either weak \citep{Lithwick2012,Batygin2013} or moderate \citep{Goldreich2014,Deck2015} eccentricity damping by the gas disc. To date, however, none of these have been shown to be universally applicable. Here we present high-resolution 2-D hydrodynamic simulations of pairs of planets undergoing convergent Type I migration into, and subsequently out of, the 2:1 MMR. We briefly review the theory of Type I migration (in Section \ref{sec:type1}), before describing our numerical method in Section \ref{sec:method} and presenting the results of our hydrodynamic simulations in Section \ref{sec:results}.  In Section \ref{sec:discussion} we discuss the physical processes responsible for breaking resonances in these calculations, and the implications of our results for the formation of compact planetary systems, before summarizing our results in Section \ref{sec:summary}.

\section{Type I migration}\label{sec:type1}
Type I migration occurs when a planet embedded in a gaseous disc is of sufficiently low mass that it does not significantly perturb the structure of the disc around it \citep[see e.g.,][for a recent review]{Baruteau2014}. In this case the region around the planet's orbit remains rich in gas, and the gravitational interaction with this gas (as well as gas further away from the planet) leads to exchange of angular momentum between planet and disc. In a linearised analysis of the waves excited in a disc by a planet, \cite{Tanaka2002} \citep[see also][]{Korycansky1993} showed that the Type I migration torque scales as
\begin{equation} \label{eq:gam0}
\Gamma_0 = \frac{q^2}{h^2}  \Sigma_p  a^4  \Omega_p^2 ,
\end{equation}
where $q$ is the planet-to-star mass ratio, $h$ is the aspect ratio of the disc, $a$ is the semi-major axis of the planet and $\Omega_p$ is the local Keplerian orbital frequency at the location of the planet. The $q^2$ term here is particularly important in the case of multiple migrating planets, since it means that for identical disc conditions, a higher mass planet will migrate proportionally faster than a lower mass one since the migration time-scale decreases linearly with mass. The orbits of mass-ordered planets therefore converge as the planets migrate, and this inevitably leads to trapping of such planets into MMRs \citep[e.g.,][]{Cresswell2006,Hands2014}.

Previous work has established that the total migration torque experienced by a low mass planet embedded in a disc is actually due to the combination of two types of torques: the Lindblad (wave) torque, and the corotation torque. Spiral density waves are launched from the Lindblad resonances both inside and outside the planet's orbit, and the difference between the outer and inner Lindblad torques drives net migration that is usually inwards. The corotation torque is generated by material that is - on average - co-orbital with the planet, and is rather sensitive to the assumptions made regarding diffusion, thermodynamics and surface-density profile in a given disc model. 

The corotation torque can take either linear \citep{Goldreich1979} or non-linear \citep{Ward1991} forms, with the non-linear form (commonly referred to as “horseshoe drag”) being the stronger of the two. The linear version of the torque occurs in the case where the planetary mass is low, and thermal/viscous diffusion in the disc is high. The non-linear torque is most readily understood as result of gas executing “horseshoe orbits” in the vicinity of the planet, whereby material approaching the planet in its rotating frame is pulled into either a higher or lower orbit by the gravity of the planet. Conservation of the fluid quantities advected in this process is the key to understanding the nature of the torque. Both the linear and non-linear corotation torques can be considered as the sum of a barotropic or vortensity-related component - generated in the presence of radial vortensity gradients in the disc - and an entropy-related component, which arises as a result of radial entropy gradients. Both of these components, in the linear and non-linear regimes, are analysed in some detail by \cite{Paardekooper2010}, producing torque formulae for all regimes. \cite{Paardekooper2011} extend these formulae to include the effects of saturation. Saturation occurs when phase mixing causes vortensity and entropy become homogeneously distributed across the horseshoe region \citep[see e.g.,][]{Masset2010}.

In the globally isothermal/barotropic case, only the vortensity-related torque acts on the planet. Vortensity is conserved along streamlines and away from shocks, and so fluid pulled across the horseshoe region by the planet must change its surface density in order to preserve its vortensity. This perturbation in surface density is what torques the planet. An investigation by \cite{Casoli2009} found that evanescent waves driven by these perturbations can actually make the corotation region asymmetric, but that this does not affect the magnitude of the horseshoe drag. In the adiabatic/non-barotropic case, vortensity is not conserved but entropy is, and the contact discontinuity generated by the advection of entropy across the corotation region generates vorticity sheets which in turn torque the planet, leading to the so-called entropy-related horseshoe drag \citep[see e.g.,][]{Masset2010}. In the locally-isothermal situation that we consider here, infinitely efficient thermal diffusion is implicit, driving the entropy related component of the horseshoe drag into the linear regime \citep{Paardekooper2010,Paardekooper2011}. This means the total horseshoe torque is the sum of this linear entropy-related torque and the non-linear vortensity-related torque from the globally-isothermal/barotropic case. Note that \cite{Masset2009,Casoli2009} identify a third “temperature-related” component of the torque, with \cite{Casoli2009} finding that the temperature-related torque is the result of vortensity that is generated efficiently in the vicinity of the planet. \cite{Paardekooper2010} separately identify this third component and also find that it arises from the source term in the vorticity equation, but note that it is negligible for all but the strongest radial temperature gradients.

By increasing the viscosity and reducing the planet mass sufficiently, one can in principle also push the vortensity-related torque into the linear regime, as in the 3D, locally isothermal simulations of \cite{Dangelo2010}. Increasing the viscosity in this way enforces the unperturbed radial vortensity profile such that vortensity is not materially conserved along streamlines. \cite{BaruteauLin2010} provide an inequality for determining the range of disc viscosities in which the barotropic horseshoe torque is both unsaturated and in the non-linear regime: 
\begin{equation}
0.16 \frac{q^{3/2}}{h^{7/2}} < \alpha < 0.16 \frac{q^{3/2}}{h^{9/2}}.
\end{equation}
For the planets we consider along and our chosen disc model, the barotropic component of the horseshoe drag should peak at $\alpha \simeq 10^{-3}$. $\alpha <  5 \times 10^{-4}$ should cause the horseshoe region to saturate, leaving only the entropy/temperature related part of the horseshoe torque. For our disc model, the barotropic component of the torque promotes outward migration, thus planets with a saturated barotropic component will experience faster inward migration driven by the Lindblad torque. $\alpha > 1 \times 10^{-2}$ would recover the linear torque as in \cite{Dangelo2010}, however, we do not consider discs of such high viscosity in this work. We thus suggest that the unsaturated torques in our simulations are most accurately described by equation 49 in \cite{Paardekooper2010}. Evaluated for our simulation setup (described below), we find
\begin{equation}
\label{eq:DangeloTotal}
\Gamma_{tot}/\Gamma_0=-0.85 - 0.9 B - A.
\end{equation}
Despite the thermodynamic complexities of the problem, this fitting formula relies only on the quantity $\Gamma_0$ (equation \ref{eq:gam0}), and the local surface density and temperature gradients in the disc (defined such that $\Sigma \propto R^{-A}$ and $T \propto R^{-B}$). We use this formula as a reference point for both testing and understanding our 2-D simulations of Type I migration. We note however that the the corotation torque almost certainly saturates for our lower viscosity runs, which would cause the overall torque to become more negative.


\section{Numerical method}\label{sec:method}
Our hydrodynamical simulations are performed using the hydrodynamical code \texttt{PLUTO} \citep{Mignone2007}, a shock-capturing Eulerian grid code. We use the standard, fixed-grid version of \texttt{PLUTO}, solving only the hydrodynamical equations. In order to minimise the computational expense of these simulations we perform them in two dimensions, using a cylindrical ($R$,$\phi$) grid. The justification for this two-dimensional set-up and some corrections made to match three-dimensional simulations are discussed below.

In order to model planet-planet and planet-disc interactions in \texttt{PLUTO}, we have coupled \texttt{PLUTO} to an $N$-body code with adaptive time-stepping. We use a 4th order Runge-Kutta integrator \citep[see e.g.,][]{Press1992} to integrate the planetary orbits and apply the force from the disc on the planets as an additional force within this integrator. We require the $N$-body code to use a time-step that is an integer fraction of the current \texttt{PLUTO} time-step, such that the two remain in sync.

In addition to the \texttt{PLUTO} simulations described in this section, as a code test we also performed an equivalent simulation with the publicly available \texttt{FARGO-3D} code \citet{Benitez2016}. Details of this simulation are contained in appendix \ref{sec:fargo}.

\subsection{Disc model} \label{sec:disc}
We initialise the disc in an axisymmetric fashion, with the surface density $\Sigma$ varying as
\begin{equation}\label{eq:r_law}
\Sigma(R) = \Sigma_{1\mathrm{AU}} \left( \frac{R}{1 \mathrm{AU}} \right)^{-1},
\end{equation}
where $\Sigma_{1\mathrm{AU}}$ is the reference surface density at 1AU. We use $\Sigma_{1\mathrm{AU}} = 1000 \mathrm{g/cm^2}$ as the fiducial value for all simulations presented here.
The pressure scale height of the disc is similarly set using a power-law:
\begin{equation} \label{eq:h_law}
H(R) = \frac{c_s(R)}{\Omega_k} = H_{1\mathrm{AU}} \cdot  \left( \frac{R}{1 \mathrm{AU}} \right)^q
\end{equation}
where we assume that $q=5/4$, leading to a disc profile that flares moderately with increasing $R$ \citep[e.g.,][]{Kenyon1987}. $H_{1\mathrm{AU}}$ is the reference value at 1AU, set to $H_{1\mathrm{AU}} = 0.05 \mathrm{AU}$. We adopt a locally isothermal equation of state such that the disc sound speed profile $c_s(R)$ is constant in time. The temperature $T(R) \propto c_s(R)^2$, and the resulting temperature power-law is therefore
\begin{equation}
T(R) \propto R^{2q-3}
\end{equation}
which for $q=5/4$ implies $T \propto R^{-1/2}$. In the nomenclature of equation \ref{eq:DangeloTotal} we therefore have have $A = 1$ and $B = 1/2$, implying $\Gamma_{tot} = -2.3 \Gamma_0$ (where $\Gamma_0$ is the scaling torque given by Equation \ref{eq:gam0}).  The negative sign here implies that we would expect each planet to lose angular momentum at every point in the disc, so we expect inward migration with a time-scale that varies depending upon local disc properties.

We note that in \cite{Hands2014}, we found that migration time-scales between $10^{3.5}$ and $10^{5.5}$ years readily produce compact systems, and that a time-scale of $\tau_{mig} \approx 10^{4.5}$yr corresponds to a disc of  $\Sigma_{1\mathrm{AU}} = 1000 \mathrm{g/cm^2}$ with $H_{1\mathrm{AU}} = 0.05$. The disc model here therefore represents a sort of ``median'' of the parametrized models in our previous work, and is thus a disc in which we would expect the assembly of compact systems to be possible. 

In order to maintain radial pressure balance we set the initial orbital velocity to be slightly sub-Keplerian. For a disc with power law density ($\rho \propto R^{-A}$) and temperature ($T \propto R^{-B}$) profiles, the required azimuthal velocity is
\begin{equation}\label{eq:densitycorrection}
\mathrm{v}_{\phi} = \mathrm{v}_{k} \cdot \left[ 1 - (A + B) \left( \frac{c_s}{\mathrm{v}_k} \right)^2 \right]^{1/2}.
\end{equation}
where $\mathrm{v}_k = \sqrt{GM/R}$ is the Keplerian orbital velocity in the mid-plane \citep[see e.g.,][]{Lodato2007}. We apply this correction to the initial conditions in all of our simulations.

We add physical angular momentum transport to our simulations through the standard $\alpha$ prescription \citep{Shakura1973}, where the viscosity is given by
\begin{equation}
\nu(R) = \alpha c_{\mathrm s} H \, ,
\end{equation}
and the dimensionless parameter $\alpha$ represents the efficiency of angular momentum transport. In addition we also observe a small amount of artificial angular momentum transport due to the numerical dissipation inherent to any fixed-grid scheme. We use a standard ring-spreading test \citep[e.g.,][]{Pringle1981,Dunhill2013a} to measure the magnitude of this numerical viscosity (using the setup described in Section \ref{sec:setup} below), and find it to be $\alpha_{\mathrm {num}} \simeq 10^{-5}$.

\subsection{Resolution and grid setup}\label{sec:setup}
In order to avoid unwanted numerical dissipation we require that the grid cells in the simulation are approximately square. That is to say, $\Delta R = R \Delta \phi$ for each and every cell. We ensure this by using a grid that is logarithmically-spaced in $R$ and linearly-spaced in $\phi$. Preliminary tests and previous work \citep[e.g.,][]{Paardekooper2013} suggest that of order 10 cells per disc scale height $H$ are required to ensure that Type I migration torques are properly resolved. The actual scale that one needs to resolve in order to correctly capture type I migration is that of the planet's corotation region, the half-width of which is given by $x_s \simeq 1.2 a \sqrt{q/h}$ \citep[see e.g.,][]{Masset2006a,Masset2016}. For our setup, there is only approximately a 10-30\% difference between the total width of the corotation region and the disc scale height for both planets, so resolving the scale height into 10 cells is sufficient. For the four models presented here, we use a grid that spans $R=[0.25$au,$3.1$au$]$ and $\phi=[0,2 \pi$]. The grid is split into 504 logarithmically spaced cells in the $R$-direction, and 1256 uniformly-spaced cells in the $\phi$-direction. As $H/R$ is not constant our numerical resolution varies slightly with radius: the grid has approximately $7.5$ cells per disc scale-height at 0.3 AU, $10$ cells per disc scale-height at 1 AU, and $12.5$ cells per disc scale-height at 2.5 AU. This ensures that the corotation radius of both planets is well resolved throughout the domain.

\subsection{Boundary conditions \& wave damping} \label{sec:bound}
We adopt zero-torque boundary conditions for the two radial edges of the computational domain, with the boundary in the $\phi$ direction being periodic. In order to combat spurious reflections of spiral density waves from the inner and outer grid boundaries (as well as other boundary effects), we adopt the standard wave-damping prescription suggested by \cite{Borro2006}. In this prescription, the following equation is integrated at each time-step in each cell belonging to two boundary regions:
\begin{equation} \label{eq:wavedamp}
\frac{\mathrm{d}x}{\mathrm{d}t} = - \frac{x - x_0}{\tau} X(R)
\end{equation}
where $x$ represents the surface density $\Sigma$ and the two velocity components in each cell, $\mathrm{v}_\phi$ and $\mathrm{v}_R$. $X(R)$ is a parabolic function that is 1 at the domain boundary and 0 at the boundary of the wave-damping zones. This ensures that waves are damped both smoothly and more strongly as they approach the simulation boundaries, without having too much affect on the rest of the computational domain. We use a simple quadratic for the function $X(R)$. The values of $\Sigma_0$, $\mathrm{v}_{\phi0}$ and $\mathrm{v}_{R0}$ provide the ``default'' values to which the damping attempts to restore each variable in each cell, which we chose to be 0, $\mathrm{v}_{k}$ and $0$ respectively  \footnote{Note that this differs from the default boundary condition in the \texttt{FARGO-3D} \citep{Benitez2016} code, which instead uses an extrapolation of the initial density profile. We discuss the implications of our boundary conditions in Appendix \ref{sec:fargo}}. Here, $\mathrm{v}_{k}$ is the Keplerian circular orbital velocity. The combination of the quadratic term in equation \ref{eq:wavedamp} and $\Sigma_0 = 0$ means that the disc settles into a state where the surface density tapers smoothly off to 0 in the wave-damping regions, which serves to shield the rest of the computational domain from the zero-torque boundary condition. As mentioned by \cite{Borro2006}, this approach clearly does not conserve mass or angular momentum, though in practice we found the disruption from this effect to be minimal compared to the spurious waves reflecting off the disc boundaries. 

We chose an inner damping region that extends from the inner boundary of the grid at $R = 0.25$au to $R = 0.28$au. The outer damping region extends from $R = 2.8$au out to the outer boundary $R = 3.1$au. The final free parameter here is the wave-damping time-scale $\tau$. we set this time-scale to
\begin{equation}
\tau = \frac{R}{\kappa c_s(R)},
\end{equation}
which is a measure of the time taken for sound waves to propagate out to radius $R$ in the disc. $\kappa$ is a constant, the exact value of which does not greatly affect the results. For the simulation set-up described here, $\kappa = 5$ was found to minimise noise from the boundaries without removing too much mass from the simulation.

\subsection{Gravitational Softening}
It has been shown that in order to reproduce migration rates from 3D simulations in 2D, the gravitational potential of the embedded planet must be softened \citep[e.g.,][]{Borro2006}. This involves modifying the Newtonian potential and acceleration  such as to avoid the singularity at $r = 0$, and is done here following the standard Plummer softening technique, in which each point-mass particle is represented as a Plummer sphere \citep[see e.g.,][]{Dehnen2011} of radius $\epsilon$, a small parameter known as the softening length.

This modification is physically motivated: a real planet is not a point mass, and gas can approach no closer than the planet's surface. More importantly in this case, however, the vertical averaging intrinsic to 2D simulations means that gas which would otherwise be vertically separated from the planet by some portion of the disc scale height $H$ is instead co-planar with the planet, and therefore able to exert a stronger force than it would in 3D. It is therefore necessary to soften the potential of the planet on a scale of order $H$ in 2D simulations.

Selecting the correct value of $\epsilon$ must be done in a way that closely matches the results of 3D simulations. Various authors have used different, albeit similar values to achieve this. For instance, \cite{Borro2006} and \cite{Paardekooper2013} used $\epsilon = 0.6 H_p$, where $H_p$ is the scale-height of the disc at the location of the planet, while \cite{Fendyke2014} used $\epsilon = 0.4 H_p$. Smaller smoothing lengths generally lead to stronger torques and faster migration. For calibration we have compared the results of our models to Equation \ref{eq:DangeloTotal}, derived by \cite{Paardekooper2010} for 3D, locally isothermal discs, which matches our 2D setup closely. For single-planet tests the migration rate scales exactly as expected with the various parameters in Equation \ref{eq:gam0}, and after calibrating against Equation \ref{eq:DangeloTotal} we adopt $\epsilon = 0.4 H_p$ in all our simulations. 

\begin{figure}
\begin{center}
\includegraphics[width=0.99\linewidth]{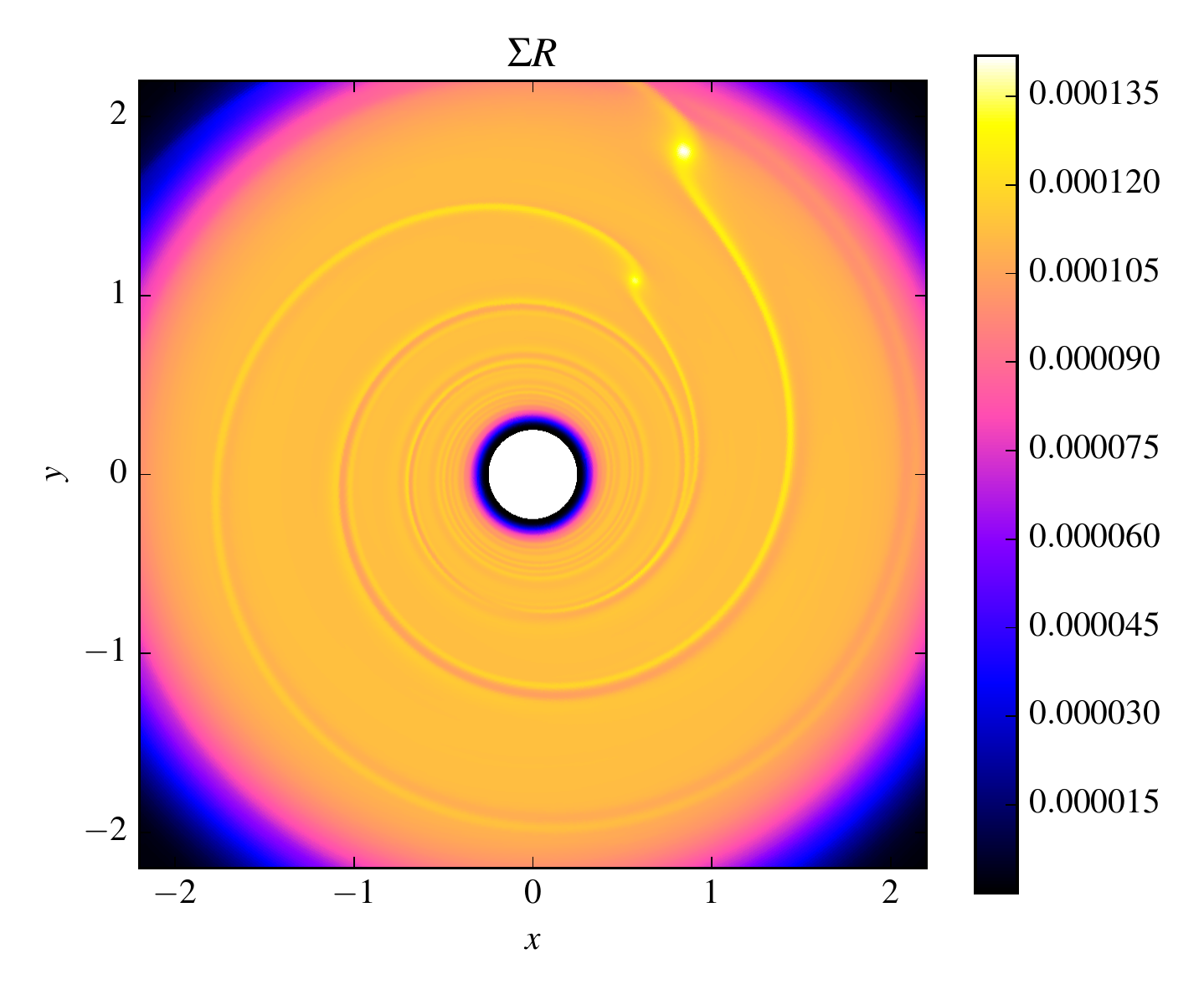}

\caption{Gas surface density from a typical snapshot of our simulations, specifically the model with $\alpha=10^{-3}$ plotted at $t=110$\,yr. To enhance the colour contrast we plot $\Sigma R$ rather than just $\Sigma$, as our disc models have $\Sigma \propto R^{-1}$. Note also that the colour scale is linear, not logarithmic. The spiral density waves induced by the planets are clearly visible, as are the (black) wave-damping zones near the grid boundaries.
\label{fig:sigma_r}}
  \end{center}
\end{figure}

\subsection{Initial conditions}
Each simulation we perform begins with two, fully formed super-Earth mass planets of masses 5 and 10 $\mathrm{M_\oplus}$ at orbital radii of 1.23 and 2.0au respectively. The more massive outer planet migrates more rapidly, and the orbits of the two planets converge. The initial spacing is chosen such that the planets begin each simulation just outside of the 2:1 MMR (with period ratio 2.075). The planetary orbits are initially circular, and have zero inclination.

We run four simulations, varying between them only the value of the $\alpha$ viscosity parameter. Our four values of $\alpha$ are 0, $1 \times 10^{-5}$, $1 \times 10^{-4}$, and $1 \times 10^{-3}$. Note that the $\alpha = 0$ case is not completely inviscid, but instead dominated by the small numerical viscosity mentioned in section \ref{sec:disc}. We run each simulation until the two planets have migrated through the 2:1 resonance, which generally takes around 3000 yr of simulation time. The simulations were run on the UK's DiRAC\footnote{\texttt{www.dirac.ac.uk}} HPC facility, specifically the Darwin Data Analytic system at the University of Cambridge, using up to 512 parallel cores.


\section{Results}\label{sec:results}

\begin{figure}
\begin{center}
\includegraphics[width=0.99\linewidth]{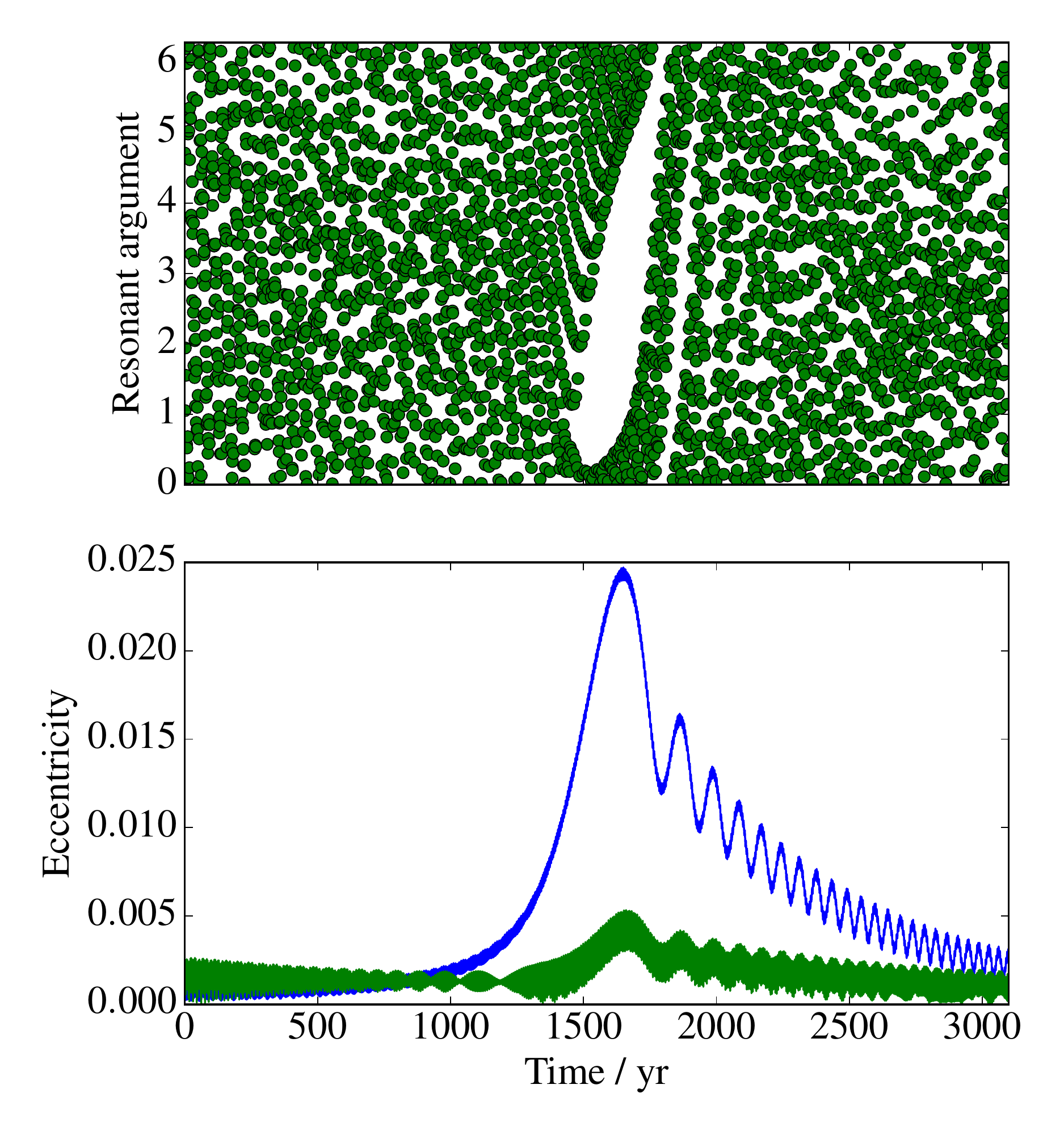}

\caption{Time evolution of the planets' resonant argument (top) and eccentricities (bottom), for the simulation with $\alpha = 10^{-3}$. Our simulations have an output frequency of $\Delta t=1$\,yr, and in the upper panel we plot one point for each output snapshot. In the lower panel the blue line denotes the eccentricity of the lower mass (5M$_{\oplus}$) inner planet, while the green line denotes the more massive (10M$_{\oplus}$) outer planet.  At $t \sim 1300$yr the resonant angle begins to librate instead of circulating, indicating that the planets have become trapped in the 2:1 MMR. The resonant torque rapidly increases the planets' eccentricity, and the lower-mass inner planet reaches $e \simeq 0.025$ within $\simeq 300$ orbits. At this point the resonance breaks, after which conventional convergent Type I migration resumes, and and the eccentricities of both planets are exponentially damped.
\label{fig:1e-3}}
  \end{center}
\end{figure}

\begin{figure}
\begin{center}
\includegraphics[width=0.99\linewidth]{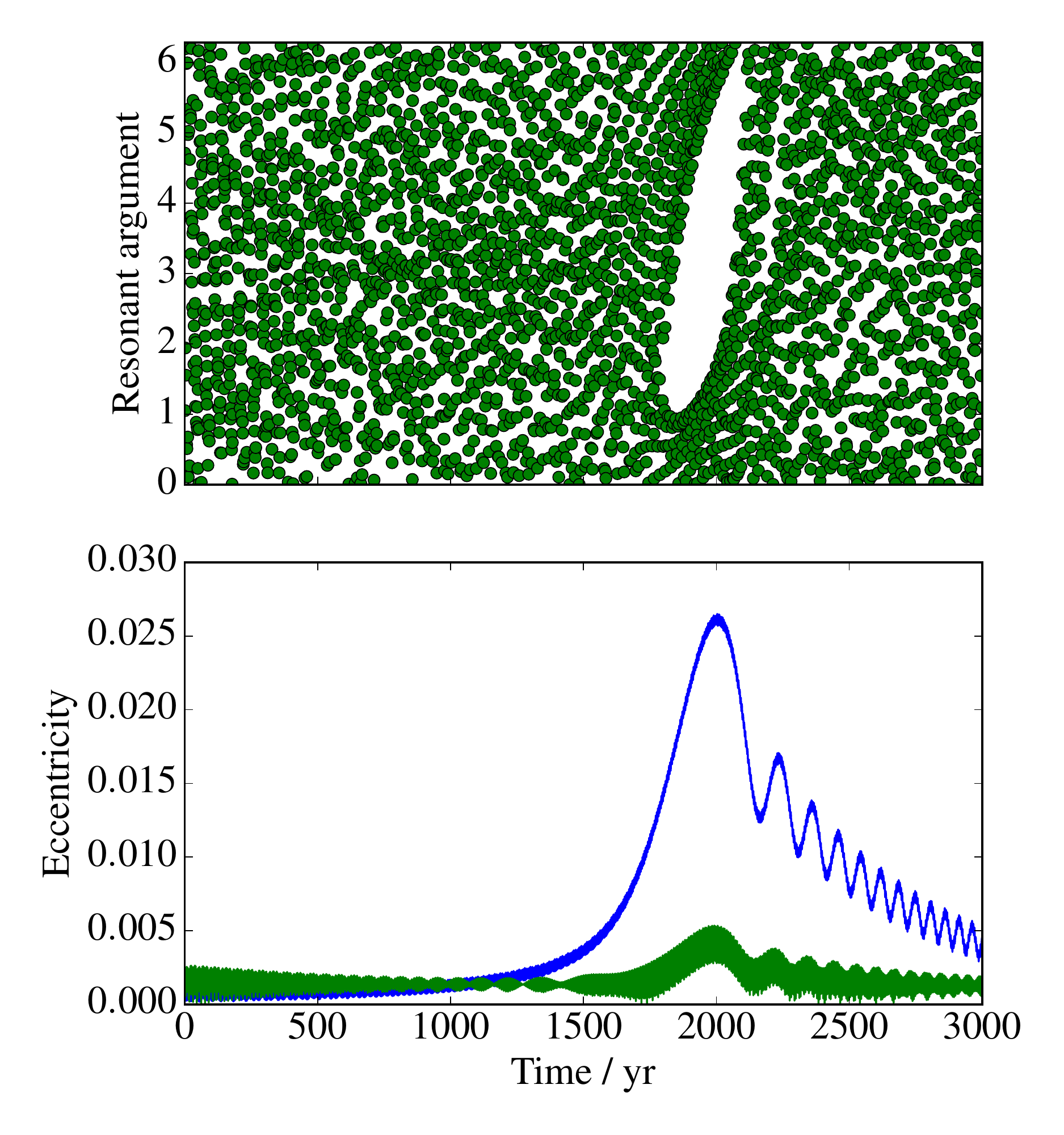}

\caption{As Fig.\ref{fig:1e-3}, but for the simulation with $\alpha = 10^{-4}$.  Compared to the $\alpha = 10^{-3}$ case the planets take longer to reach the 2:1 resonance, the peak eccentricity is slightly higher, and the resonance persists somewhat longer before breaking, but the qualitative behaviour is identical.\label{fig:1e-4}}
  \end{center}
\end{figure}

\begin{figure}
\begin{center}
\includegraphics[width=0.99\linewidth]{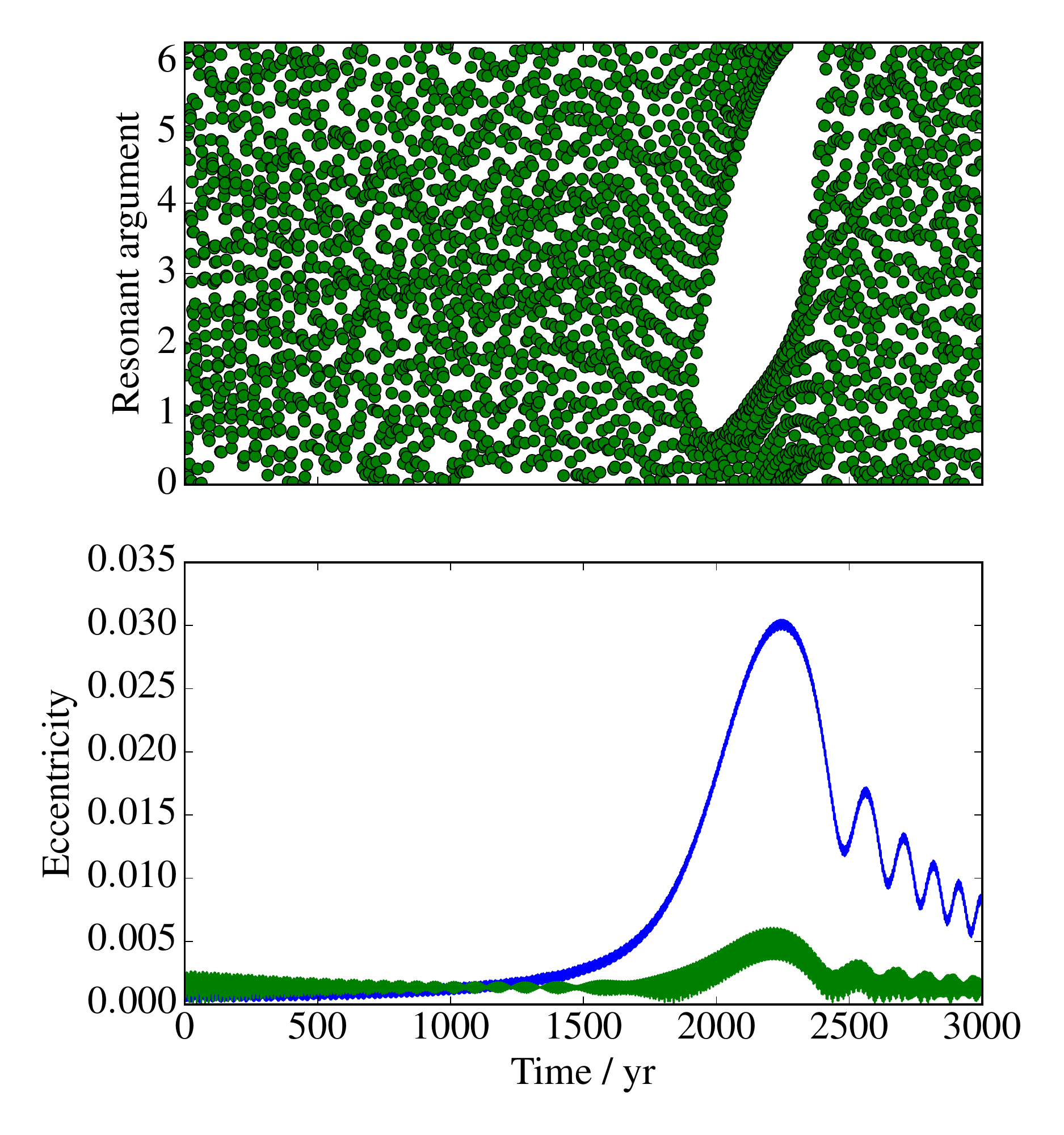}

\caption{As Fig.\ref{fig:1e-3}, but for the inviscid simulation.  Again the details differ slightly, and the resonance survives for almost twice as long as in the $\alpha = 10^{-3}$ simulation, but the qualitative behaviour is apparently independent of the magnitude of the viscosity.\label{fig:inviscid}}
  \end{center}
\end{figure}

\begin{figure}
\begin{center}
\includegraphics[width=0.99\linewidth]{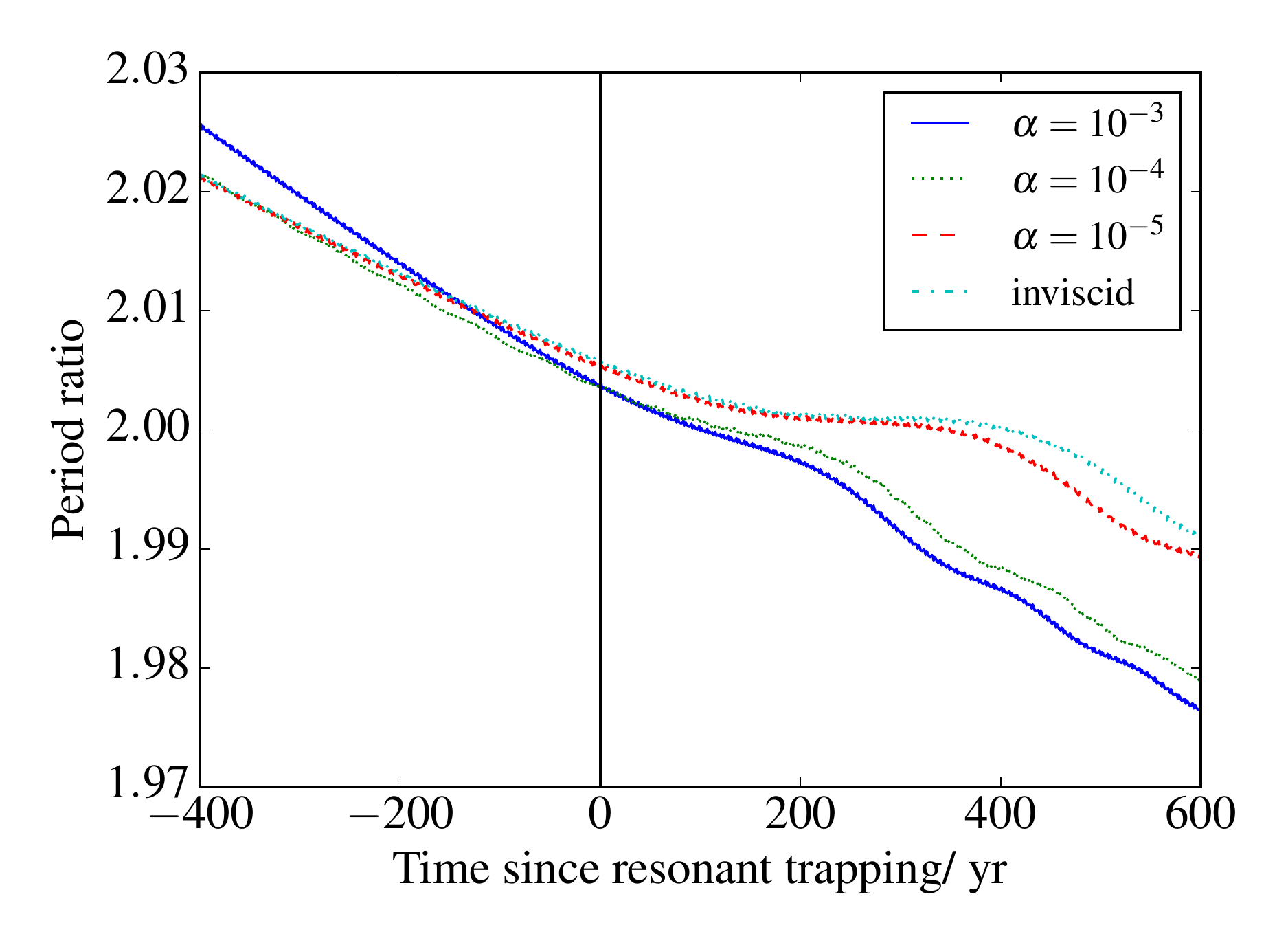}

\caption{Evolution of the planets' period ratio in each of our simulations. The time axis is normalised to the point where each simulation first becomes trapped in resonance. In each simulation the resonance persists for a few hundred orbits before the planets escape and continue convergent migration, but the lifetime of the resonance decreases significantly with increasing disc viscosity. 
\label{fig:combinedperiodplots}}
  \end{center}
\end{figure}

Fig.\,\ref{fig:sigma_r} shows the gas density from a snapshot in the simulation with $\alpha =10^{-3}$: the spiral density waves induced by both planets are clearly seen, as are the changes in density in the wave-damping zones near the grid boundaries. The evolution of the planets' orbits in the simulations with $\alpha =10^{-3}$, $\alpha =10^{-4}$ and the inviscid disc are shown in Figs.\,\ref{fig:1e-3}, \ref{fig:1e-4} and \ref{fig:inviscid}, respectively, while the evolution of the planets' period ratio in all four simulations is shown in Fig.\,\ref{fig:combinedperiodplots}. (The simulation with $\alpha =10^{-5}$ is almost identical to the inviscid model, so for clarity we do not plot its orbital evolution here.) The qualitative behaviour is the same in all four simulations. In all cases both planets initially migrate steadily inwards, with the more massive outer planet migrating faster than the less massive inner planet, as expected from Equation \ref{eq:gam0}. As the planets' period ratio approaches 2 they become trapped in resonance, and we see the resonant argument begin to librate instead of circulating. At this point the eccentricities of the two planets rise rapidly, reaching $e \simeq 0.03$ for the less massive inner planet. Note that despite the relatively small mass ratio of the planets, the inner planet attains a significantly lower eccentricity than the more massive, outer planet. There are several reasons for this. First, the resonances break before an equilibrium eccentricity is reached, implying that the eccentricities of both planets would otherwise continue to grow and possibly address this imbalance. Furthermore, if one follows the disturbing function approach to analysing the resonant interaction, one finds that the direct and indirect contributions to the disturbing potential for the exterior planet in the 2:1 resonance almost cancel to first order in eccentricity \citep[see e.g.,][]{Teyssandier2014}, leading to little evolution for the exterior planet. The eccentricity pumping in the resonance continues for a few hundred orbits in each case, with the period ratio remaining approximately constant (as one would expect for planets trapped in resonance). The eccentricity then declines, and at the same time the resonant argument ceases librating and the planets begin converging once again, indicating that the planets have escaped from the 2:1 resonance. We expect a simple resonance like this to be stable, with only mild, pendulum-like oscillations in eccentricity and inclination \citep[see e.g.,][]{Davis2014}, so the natural interpretation is that the MMRs in our simulations are broken by the planet-disc torques.

To confirm that this resonance-breaking is the result of hydrodynamic effects, as opposed to secular evolution or $N$-body integration issues, we perform an additional simulation using only the \emph{N}-body (gravitational) integrator. We start this calculation from the state of the $N$-body particles in the inviscid simulation at $t = 2200$yr, at which point the two planets are trapped in the 2:1 resonance (see Fig.\ref{fig:inviscid}). We then integrate this configuration forwards using the same adaptive time-step Runge-Kutta method, but without the gas disc or any hydrodynamics, until $t=10,000$yr. The planets remain locked in resonance throughout, with the eccentricity of the inner planet oscillating between $e \simeq 0.02$--0.045 on a period of $\sim 1000$yr. This test demonstrates clearly that in the absence of the gas disc this resonance is stable over very long time-scales ($>10^4$yr). We therefore conclude that the disc-planet interactions in our simulations act to break the 2:1 MMR between the planets, and that this mechanism operates robustly over a wide range in $\alpha$.

Fig.\,\ref{fig:combinedperiodplots} shows the evolution of the period ratio between the two planets for each of the simulations discussed above. The origin of the time axis in this figure is normalised to the time at which the two planets first become trapped in resonance (measured by eye as the point at which the resonant argument begins to librate). In each simulation the period ratio between the two planet decreases monotonically until it reaches a value slightly greater than 2.0, then remains approximately constant while the planets are trapped in resonance. Once the resonance is broken convergent migration resumes and the period ratio declines once more. However, Fig.\,\ref{fig:combinedperiodplots} also shows that the lifetime of the resonance depends on the disc viscosity, with higher viscosity parameters resulting in faster breaking of the resonance. We see that the 2:1 resonance resonance persists for more than twice as long in the inviscid disc as in the $\alpha = 10^{-3}$ case, despite the fact that accretion has significantly depleted the disc surface density (by 30-35\%) in the $\alpha = 10^{-3}$ simulation (see Table \ref{tab:timescales}).

\subsection{Time-scales}

\begin{table}
\centering

\begin{tabular}{lllll} 
\emph{Run}: & inviscid & $\alpha = 1 \times 10^{-5}$  & $\alpha = 1 \times 10^{-4}$ & $\alpha = 1 \times 10^{-3}$ \\ \hline \hline
$t_{start}$ (yr) & 2500 & 2450 & 2200 & 1750\\ \hline
$t_{end}$ (yr)& 3000 & 2950 & 2700 & 2250\\ \hline
$a_1$ (AU) & 1.183 & 1.185 & 1.191 & 1.198\\ \hline
$a_2$ (AU) & 1.866 & 1.868 & 1.876 & 1.889\\ \hline
$\Sigma_1$ (g/cm$^2$)& 794.493 & 798.432 & 804.955 & 777.735\\ \hline
$\Sigma_2$ (g/cm$^2$)& 493.120 & 485.439 & 464.335 & 316.592\\ \hline
$(H/R)_1$ & 0.052 & 0.052 & 0.052 & 0.052\\ \hline
$(H/R)_2$ & 0.058 & 0.058 & 0.059 & 0.059\\ \hline
$\tau_{a,1}$ (yr)& 7.66$\times 10^4$ & 7.70$\times 10^4$ & 8.24$\times 10^4$ & 7.90$\times 10^4$\\ \hline
$\tau_{a,2}$ (yr)& 3.88$\times 10^4 $& 3.82$\times 10^4 $& 3.53$\times 10^4$ & 3.50$\times 10^4$\\ \hline
$\tau_{e,1}$ (yr)& 5.79$\times 10^2$ & 5.57$\times 10^2$ & 5.68$\times 10^2$ & 7.56$\times 10^2$ \\ \hline
$\tau_{e,2}$ (yr)& 9.34$\times 10^2 $& 9.20$\times 10^2 $& 8.46$\times 10^2$ & 1.22$\times 10^3 $\\ \hline
$K_{1}$ & 132.259 & 138.224 & 145.079 & 104.494\\ \hline
$K_{2}$ & 41.544 & 41.538 & 41.766 & 28.682\\ \hline
\end{tabular}
\caption{\label{tab:timescales}Table showing eccentricity damping ($\tau_{e,p}$) and migration ($\tau_{a,p}$) time-scales fitted to each of our simulations in the post-resonance phase for each planet $p=1,2$. Here, $p=1$ represents the inner, lower-mass planet and $p=2$ represents the outer, higher-mass planet. $T_{start}$ and $T_{end}$ define the period of time in the simulation over which the fits were made.  $a_p$, $\Sigma_p$ and $(H/R)_p$  represent the semi-major axis of the planet, and the disc surface density and aspect ratio at this location respectively. Finally,  $K_p$ represents the constant of proportionality between $\tau_{e,p}$ and $\tau_{a,p}$ for each planet, as defined in equation \ref{eq:Kpnew}.}
\end{table}

To compare our simulations directly to other work in the literature as well as our previous work, we fit migration and eccentricity damping time-scales to the changes of the orbital elements of each planet in each separate simulation. Here, we assume, as in \citet{Hands2014} and \citet{Hands2016}, that the semi-major axis of each planet varies as
\begin{equation}
\frac{da_p}{dt} = -\frac{a_p}{\tau_{a,p}}
\end{equation}
where $a_p$ is the semi-major axis of each planet $p$, and $\tau_{a,p}$ is the individual migration time-scale for that planet. The eccentricity $e_p$ of each planet then varies as  
\begin{equation}\label{eq:Kpnew}
\frac{de_p}{dt} = -\frac{e_p}{\tau_{e,p}} = -K_p\frac{e_p}{\tau_{a,p}},
\end{equation}
where $K_p$ is a factor that determines how much faster eccentricity damping is than migration. This prescription implies that both elements are damped exponentially, each with a characteristic e-folding time.

To fit time-scales to our simulations, we first manually select a region of time shortly after the resonance has broken (i.e., the period of time in which the resonant argument once again begins to circulate). This ensures that the period of the simulation we fit to is not affected by resonant interactions, and thus that the effect we are seeing is purely a result of planet-disc interactions. Figs.\,\ref{fig:1e-3}, \ref{fig:1e-4} and \ref{fig:inviscid} show that the moderate eccentricities are damped essentially exponentially after this point, making them ideal for fitting. This region begins at a slightly different time depending on which simulation is being fitted, but we keep the width of this region as 500\,yr for each fit. We then fit exponential curves to the semi-major axes and eccentricities of each planet over this region, using a non-linear least-squares method, to determine the time-scales ${\tau_{a,p}}$ and ${\tau_{e,p}}$, and thence $K_p$, for each planet. We also calculate the disc-aspect ratio $(H/R)_p$ and azimuthally-averaged surface density $\Sigma_p$ at the semi-major axis of each planet exactly half-way through this period. 

The results of this analysis are shown in Table \ref{tab:timescales}. The fitted migration time-scales confirm the assumption of our previous work, and Section \ref{sec:disc}, by showing that for our disc model, $\tau_{a,p} \simeq 10^{4.5} \mathrm{yr}$. The higher-mass planet migrates almost twice as fast as the lower-mass inner planet, as expected (see Section \ref{sec:type1}). This analysis also confirms that the range of $K_p$ values that \cite{Hands2014} found could form planetary systems ($10^{1.5} \leq K \leq 10^{2.5}$) is reasonable for the disc model presented here, and that this disc model does indeed represent a ``median'' of the parametrized models in our previous work.

The value of $K_p$ is naturally an important constraint for models of disc-driven migration. Previous analytical work by \cite{Tanaka2004} suggests that the value of $K_p$ scales as
\begin{equation}\label{eq:estK}
K_p \propto \left( \frac{H_p}{R_p} \right) ^2
\end{equation}
where $H_p$ is the scale-height of the disc at the radial location of the planet, $R_p$. This relation implies that in a moderately flaring disk such as ours, $K_p$ ought to be greater for planets further out in the disc. Our results suggest this is not the case: the value of $K_p$ for the outer, more massive planet is consistently lower than for the inner planet (by a factor $\simeq4$). The reason for this discrepancy is unclear and possibly warrants further investigation. However, we note that since the results of our \texttt{FARGO} simulations in appendix \ref{sec:fargo} show good agreement with those presented here, this is unlikely to be due to our numerical method.

\subsection{Torque analysis}
\begin{figure}
\begin{center}
\includegraphics[width=0.99\linewidth]{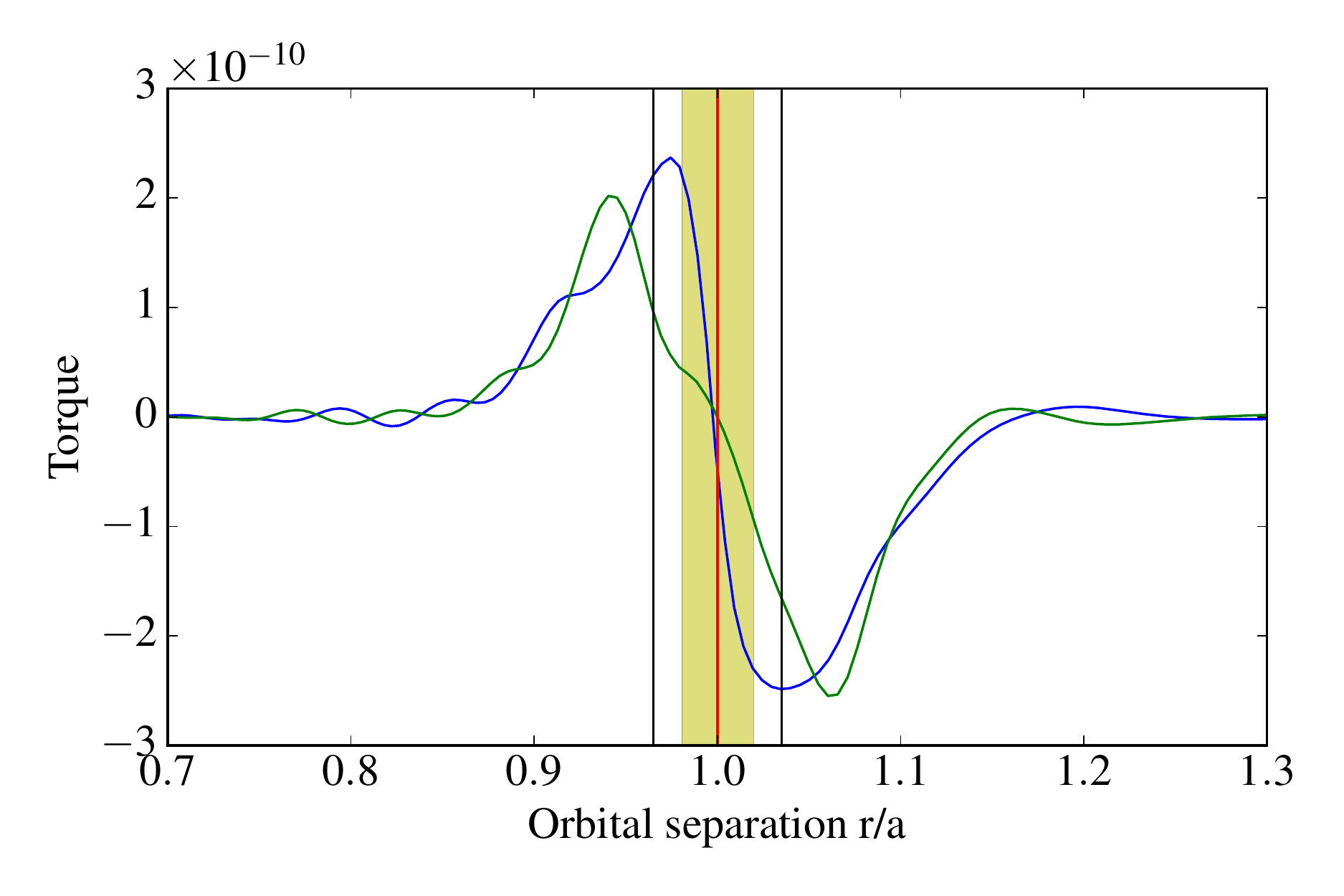}

\caption{Azimuthally-averaged disc torques on the inner planet for $\alpha=10^{-3}$ at $t=800$yr (i.e., before resonant trapping; blue curve) and $t=1600$yr (i.e., shortly before resonance breaking; green curve). The radial coordinate is normalised to the planet's semi-major axis, which is marked as a red line (to guide the eye). The black vertical lines are drawn at $\pm2H/3$ from the planet's position, at the at the location where the Lindblad resonances pile-up, while the yellow shading indicates the approximate extent of the corotation region. There is a pronounced change in the radial profile of the torque once the plants are trapped in resonance, with the peak torques arising roughly twice as far from the planet's position as they do out of resonance. We attribute this change in the torque profile to the increase in the planet's orbital eccentricity.\label{fig:torques1e-3}}
  \end{center}
\end{figure}

\begin{figure}
\begin{center}
\includegraphics[width=0.99\linewidth]{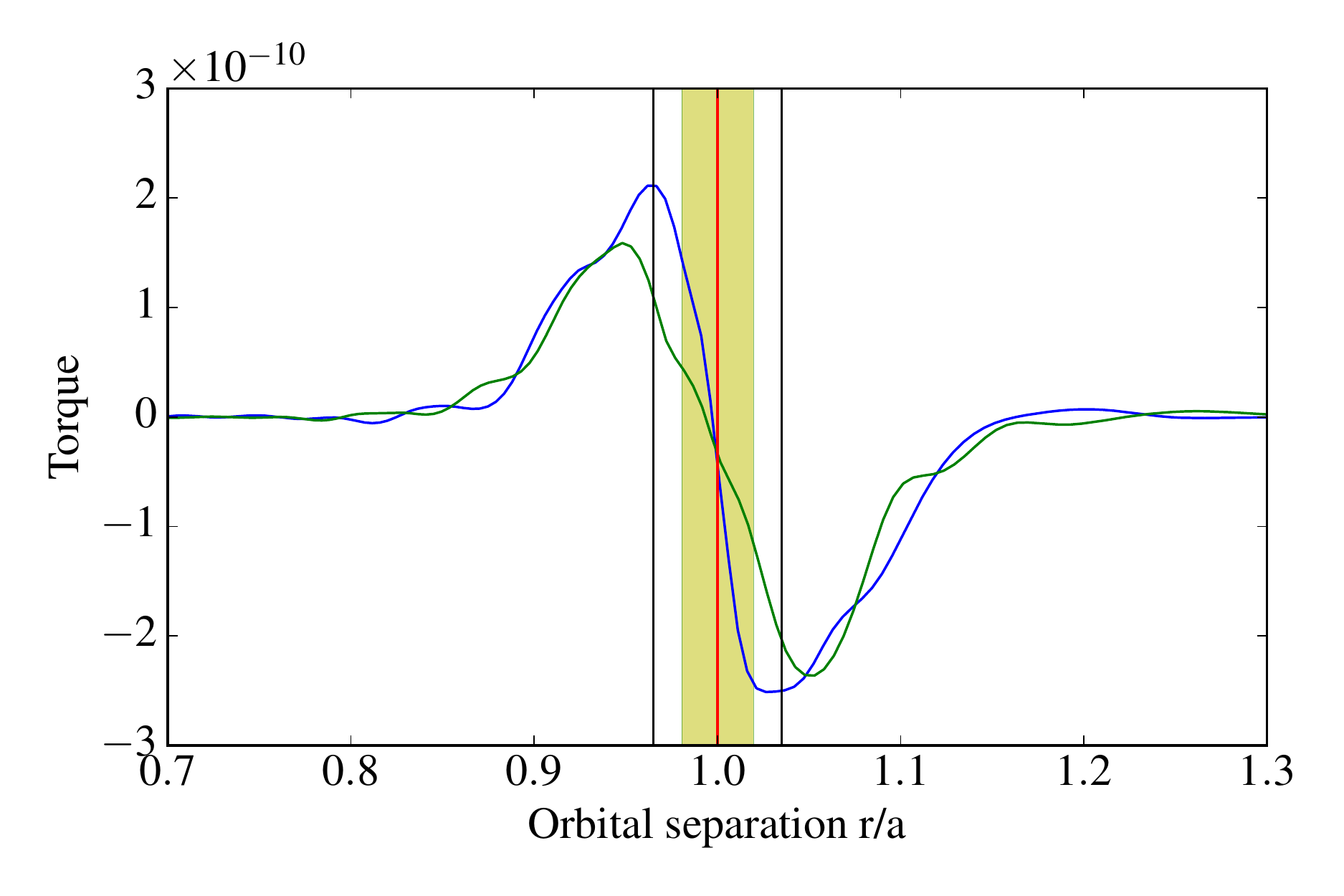}

\caption{As Fig.\,\ref{fig:torques1e-3}, but for the simulation with $\alpha=10^{-5}$. Here the disc torques on the inner planet are plotted at $t=1000$yr (i.e., before resonant trapping; blue curve) and $t=2100$yr (i.e., shortly before resonance breaking; green curve). Again we see a pronounced change in the torque profile when the planets are in resonance, with substantial differences in the corotation region and the peak torques shifted away from the planet's position.\label{fig:torques1e-5}}
  \end{center}
\end{figure}

The mechanism responsible for breaking the resonances in our simulations is not immediately clear, but there is evidently a change in the Type I migration torque during the period in which the planets are in resonance. In Fig.\,\ref{fig:torques1e-3} and \ref{fig:torques1e-5} we plot the azimuthally-averaged disc torques acting on the inner planet, as a function of radius, for $\alpha = 10^{-3}$ and $10^{-5}$. (The planet-planet torques are not included in these figures.) To reduce noise in the torques caused by fluctuations around the planet's orbit, we also average the torques over a short time window. To compute the torque we take 12 adjacent snapshots (which are generated once per year of simulation time), average the disc torques in each snapshot azimuthally, and then average these torques across all 12 snapshots. It is immediately clear from Figs.\,\ref{fig:torques1e-3} and \ref{fig:torques1e-5} that there is a pronounced change in magnitude of the torques in the corotation region. Out of resonance the torque peaks at $\simeq \pm 2H/3$ from from the planet's position, where the Lindblad resonances pile up. In resonance, however, we see significant structure in the torque profile in the corotation region, and the peak torques arise much further from the planet (approximately twice as far from the planet compared to out of resonance). This is most likely the result of changes in the torques caused by the increased eccentricity of the planets, discussed further in section \ref{sec:torquechange}. 

The disc properties do not change as the planets move into and out of resonance, so these changes in the disc-planet torque must be due to changes in the planet's orbit, specifically the increase in the orbital eccentricity. This is further supported by the fact that the disc-planet torque is observed to vary periodically with the same period as the eccentricity oscillations driven by the resonance.  We therefore conclude that the disc-planet torques change both qualitatively and quantitatively as the planets' eccentricities grow, and that these torques are responsible for breaking the MMRs in our simulations.

\section{Discussion}\label{sec:discussion}
Our simulations demonstrate that disc-planet interactions can cause migrating planets to escape from the 2:1 MMR for a wide range of $\alpha$ values, but thus far we have not identified a physical mechanism for this resonance breaking. The observed change in the form and effect of the disc-planet torque is clearly driven by the increase in the planet's orbital eccentricity when in resonance, so there are two plausible physical mechanisms to consider: i) changes in the disc-planet torques due to the excitation of additional Lindblad and/or corotation resonances when $e>0$ \citep{Papaloizou2000,Fendyke2014}; or ii) rapid eccentricity damping by the disc leading to overstable librations once the eccentricity exceeds a threshold value \citep{Goldreich2014}. We see evidence for both of these processes in our simulations, so we discuss each of them in turn.

\subsection{Changes in the disc-planet torques} \label{sec:torquechange}

It has long been recognised that the excitation of eccentric Lindblad resonances and/or non-co-orbital corotation resonances can qualitatively change the disc-planet interaction, but little attention has been paid to this effect as a mechanism for breaking resonances. \cite{Papaloizou2000} found that the net Lindblad torque can actually change sign when $e > 1.1(H/R)$. \footnote{Note, however, that the reverse torque does not necessarily drive outward migration, but may simply circularise the planet \citep{Masset2008}.} This is a result of the eccentricity of the planet allowing it to cross the comb of circular Lindblad resonances. \cite{Fendyke2014} later confirmed this effect, and further showed that the corotation torque reduces exponentially in magnitude when planets attain small eccentricities. \cite{Fendyke2014} also showed that the Lindblad torques typically decrease in magnitude before they change sign. Although our planets fall somewhat short of the suggested threshold eccentricity of $1.1(H/R)$ required for the torque to change sign [we see peak values of $e \simeq 0.8(H/R)$], the reduction in magnitude and movement of the peak of the disc torque in the vicinity of the planet in Fig.\,\ref{fig:torques1e-3} is entirely consistent with these predictions. Moreover, once eccentricity starts to grow we see large oscillations in the disc-planet torque on the planet's orbital period, and the net torque on the inner (lower mass) planet is positive for a significant fraction of each orbit. Out of resonance, when the planets have minimal eccentricities, the net torque on both planets is consistently negative. This is therefore a plausible explanation for resonance-breaking: the change in magnitude of the planet-disc torque as eccentricity grows is sufficient to push the inner planet out of resonance, and the change in sign is consistent with the inner planet escaping the resonance in the outward direction.

\subsection{Overstable librations} \label{sec:overlib}
\begin{figure}
\begin{center}
\includegraphics[width=0.99\linewidth]{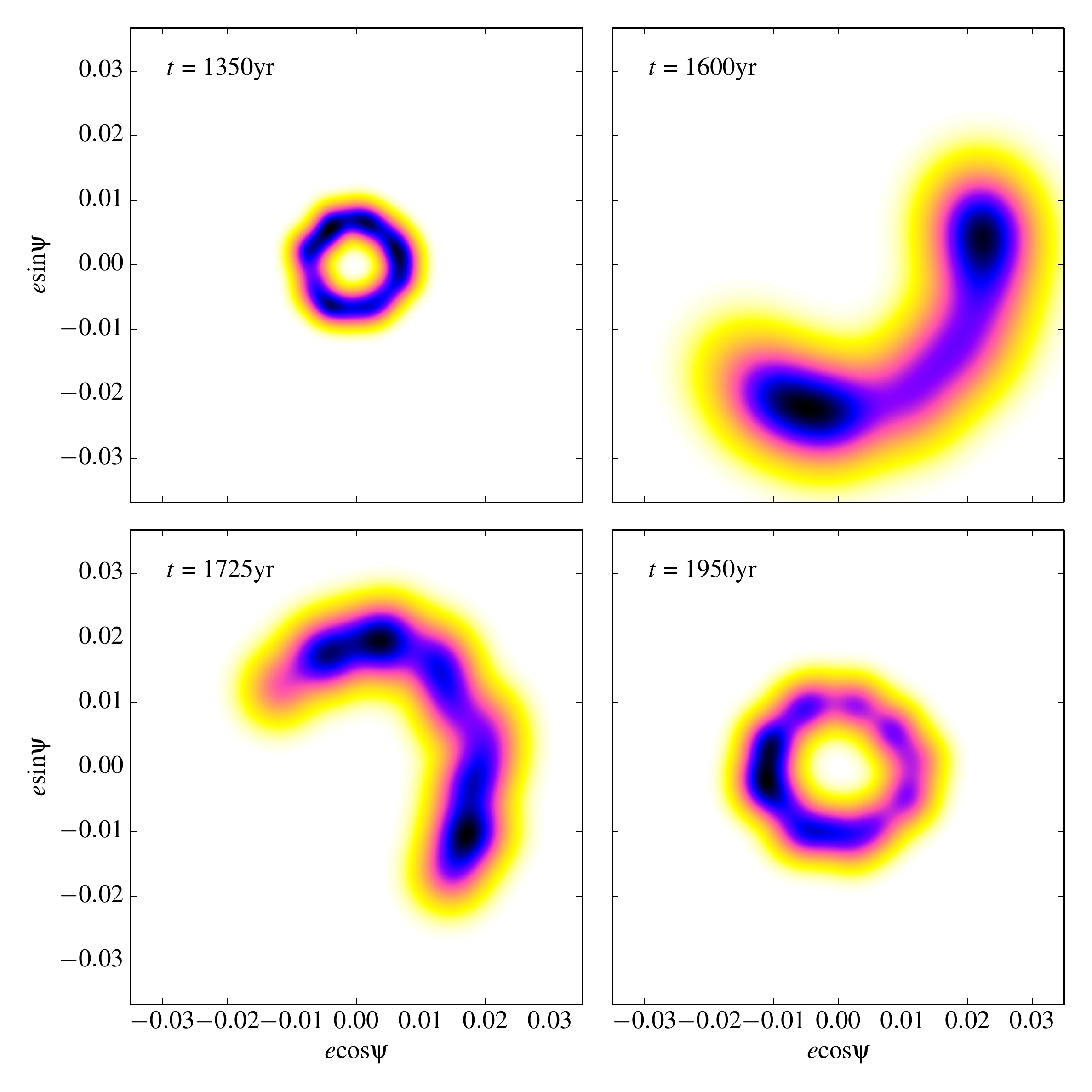}

\caption{Evolution of the simulation with $\alpha = 10^{-3}$ in the $e \sin \psi$ / $e \cos \psi$ plane (where $\psi$ is the resonant argument and $e$ is the eccentricity of the inner planet). The colour scale indicates the density of points in this plane over 50\,yr of simulation time. As we only have simulation output data at discrete intervals (one output dump per yr), we smooth the discrete output points using a two-dimensional Gaussian profile with half-width $\sigma = 0.2e$. The colour scale is normalised to the peak value in all cases. The sub-structure in these plots is an artefact of our low sampling frequency, but the coloured locus is an accurate representation the system's behaviour. We see the resonant argument move from circulation (top left) to libration (top right) as the planets enter the resonance. Once in resonance the libration amplitude increases steadily (due to overstable damping; bottom left) until the system escapes resonance and the argument once again circulates (bottom right).
\label{fig:ecossin}}
  \end{center}
\end{figure}

Alternatively, \citet[][hereafter GS14; see also \citealt{Deck2015}]{Goldreich2014} have suggested that eccentricity damping by the disc can lead to overstable librations of the resonant argument if $e \gtrsim (M_{\mathrm p}/M_*)^{1/3}$. This overstability causes the amplitude of the libration to grow rapidly, until the planets escape from resonance and their eccentricities are rapidly damped back to zero. GS14 used analytic theory and \emph{N}-body simulations to show that this mechanism operates effectively with a fixed eccentricity damping time-scale $\tau_e$ if the ratio $K_p = \tau_a/\tau_e$ is sufficiently small. However, to our knowledge this effect has not yet been seen in hydrodynamic calculations.

For the general case of the $j$+1:$j$ resonance, GS14 showed that planets can escape from resonance if the criterion 
\begin{equation}
K_p < \frac{3}{2} \frac{1}{3j+p} \left(\frac{j^2p}{8 \beta}\right)^{2/3} \left(\frac{M_*}{M_2}\right)^{2/3}\end{equation}
is satisfied\footnote{This is Equation A5 from GS14, re-arranged as a limit on $K_p$ rather than on $M_2$. The leading factor of $3/2$ arises as we choose to work with the semi-major axis damping (migration) time-scale $\tau_a$, instead of the time-scale for damping the mean motion.}. Here $\beta \simeq 0.8j$, and as before $M_2$ is the mass of the outer planet. The constant $p$ (defined in GS14's Equation 2; see also \citealt{Deck2015}) accounts for the change in a planet's mean-motion due to eccentricity damping. GS14 primarily considered the case $p\simeq 3$, which corresponds to energy dissipation at constant angular momentum, and is appropriate for star-planet tidal interactions. The dependence on $p$ is weak, but here a lower value is more appropriate. For the 2:1 MMR we have $j=1$, and for $p=3/2$ this criterion reduces to
\begin{equation}
K_p < C \left(\frac{M_*}{M_2}\right)^{2/3} \, ,
\end{equation} 
where the constant $C = 0.127$. In our simulations $M_2 = 10$\,M$_{\oplus}$, so escape from resonance due to overstable librations requires\footnote{This is not strictly correct, as in GS14's calculation the orbit of the outer planet was fixed. In the general case where both planets migrate, capture into / escape from resonance depends on the planets' relative migration time-scale, rather than $\tau_a$ for just the inner planet. However, generalising GS14's analysis to consider both planets migrating is beyond the scope of this paper, so we discuss our results in terms of $K_p$ for each planet.} $K_p<131$ and $e \gtrsim (M_1/M_*)^{1/3} = 0.0247$. 
Our simulations show peak eccentricities $e \simeq 0.025$--0.030, and our planets have $K_p \simeq 40$ \& 130 (for the outer and inner planets, respectively; see Table \ref{tab:timescales}). Our simulations therefore marginally satisfy GS14's conditions for overstability, and the qualitative evolution in Figs.\,\ref{fig:1e-3}--\ref{fig:inviscid} is very similar to that described above. We further found that repeating the \emph{N}-body-only calculation from Section \ref{sec:results} and imposing eccentricity damping on the same time-scales \citep[using the damping algorithm described in][]{Hands2014} does indeed lead to the planets escaping the 2:1 resonance in the manner predicted by GS14. The behaviour of the hydrodynamic simulations, however, is rather more complex.

To investigate this issue further, we follow GS14 (specifically their Fig.\,5) and consider the evolution in the $e \sin \psi$ / $e \cos \psi$ plane (where $\psi$ is the resonant argument and $e$ is the eccentricity of the inner planet). Fig.\,\ref{fig:ecossin} shows the evolution of our simulation with $\alpha = 10^{-3}$. We see the resonant argument moves from circulation (an approximately circular locus in the $e \sin \psi$ / $e \cos \psi$ plane) to libration (an arc spanning a limited range of angles $\psi$) as the system enters resonance. The libration amplitude stays more or less constant (at $\sim 2\pi/3$) from $t\simeq 1550$--1700\,yr as the eccentricity peaks (at $e \simeq 0.25$). From this point the eccentricity starts to decay and the amplitude of the librations increases steadily, returning to circulation as the system escapes resonance at $t\simeq 1900$\,yr.  This behaviour is broadly similar to that predicted by GS14, but there are some notable differences. In particular, we see the planet's eccentricity decay significantly as the librations grow in amplitude (i.e., before the system escapes resonance), whereas for $K_p = 50$ GS14 found that the eccentricity increased steadily as the libration amplitude increased. We also do not see the substantial eccentricity oscillations prior to escape that GS14 observed in their damped \emph{N}-body calculations. We attribute this to the difference in eccentricity damping between these two approaches. GS14 imposed a constant eccentricity damping time-scale, while in our simulations eccentricity is damped self-consistently by the disc-planet torques (with $K_p \simeq 130$ for the inner planet and $K_p \simeq 40$ for the outer planet; see Table \ref{tab:timescales}). Moreover, in the hydrodynamic calculations the disc-planet torques vary substantially around the orbit once $e\gtrsim 0.01$, so on short ($\sim$orbital) time-scales a single eccentricity damping time-scale is a poor approximation to the damping process. We therefore conclude that overstable librations driven by strong eccentricity damping are responsible for breaking the MMRs in our simulations, but caution that the behaviour of this mechanism in real discs is somewhat different to the idealised calculations of GS14.

As a final remark, we note that values of $K_p \simeq 100$ are normal for locally isothermal discs, more complex thermodynamic treatment may alter this picture. Radiative simulations such as those by \cite{Bitsch2013,Bitsch2013a} demonstrate that slow, stalled or even reversed type I migration can occur in significant portions of realistic discs. The dramatic reduction in migration rate in these regions could in principle inflate the value of $K_p$ significantly, making this mechanism unworkable in these regions and promoting longer-lived resonant trapping.

\subsection{Limitations of this work}
These results are extremely encouraging, but understanding how this mechanism behaves across a much broader range of disc conditions is necessary in order to confirm its robustness and applicability. Unfortunately, the simulations presented here are computationally expensive, and exploring a large parameter space is beyond the scope of this study. We have only considered the 2:1 MMR here, but other closer-spaced resonances must also be crossed or broken to produce the observed diversity of exoplanet architectures. However, if this mechanism can break a strong first-order MMR such 2:1, it is likely to be effective at breaking other, weaker resonances as well.

Moving further, an obvious next step would be to repeat these simulations with a lower disc surface density $\Sigma$, which would weaken the disc-planet torque; presumably there is a threshold below which the torques are too weak to break MMRs. We would also like to explore a range of planet masses, and consider the case where the more massive planet has the shorter period \citep[e.g.,][]{Deck2015}. We note, however, that the majority of compact systems are mass ordered, with larger planets exterior to smaller ones, and thus the case explored in this work is perhaps more immediately relevant. Additionally, both of these are subject to computational limitations, as lower surface densities and lower planet masses both result in longer migration time-scales (Equation \ref{eq:gam0}), which require much longer run-times in simulations. More massive planets migrate more quickly, which is computationally advantageous, but here we run into physical limitations: increasing the planet mass to $\gtrsim20$M$_{\oplus}$ pushes us beyond straightforward Type I migration, and into the more complex regime where the planets' back-reaction on the disc is no longer negligible. 

If our interpretation is correct and overstable librations are the primary resonance-breaking mechanism, then the critical parameter is the time-scale ratio $K_p$; GS14 find that overstability does not occur when $K_p\gtrsim 100$. Moving further into the overstable regime means reducing the value of $K_p$, which suggests that future work should consider thinner discs (see Equation \ref{eq:Kpnew}). However, in this regime the behaviour may be more complex, as for thinner discs we are likely to see $e > (H/R)$ (and corresponding changes in the disc-planet torque) before the threshold eccentricity for overstability is reached. Moreover, our assumption of a locally isothermal equation of state is significant simplification, particularly if (as suggested by Figs.\,\ref{fig:torques1e-3} \& \ref{fig:torques1e-5}) the corotation region contributes significantly in the disc-planet torque. Finally we note that these simulations (and the calculations of GS14) are limited to two-dimensions, and therefore neglect the possibility that MMRs can excite inclination as well as eccentricity. Three dimensional calculations seem likely to reveal additional complexity.

\subsection{Implications}
In recent years the paucity of MMRs in {\it Kepler} multi-planet systems has commonly been invoked as an argument against migration, and has led several groups to consider scenarios where even tightly-packed planets form {\it in situ} at radii $\lesssim 0.1$AU \citep[e.g.][]{Hansen2012,Chiang2013,Boley2016}. Our results, however, demonstrate that resonance-breaking may occur for planets undergoing Type I migration in realistic protoplanetary discs, and in particular in a disc model which we consider to be suitable for compact system formation following our previous work \citep{Hands2014}.  
If this mechanism applies widely, then Type I migration need not result in a high incidence of MMRs in compact multi-planet systems \citep[as predicted by studies using parametrized migration models; e.g.,][]{Hands2014,Coleman2016}. Our results show that planets can form conventionally at $\sim$AU radii (i.e., beyond the snow-line), and then migrate to radii $\ll 1$AU without becoming permanently trapped in resonance. Detailed discussion of {\it in situ} formation models is beyond the scope of this work. However, if disc-planet interactions break resonances for a wide range of disc models, then the observed architectures of compact multiple systems can be attributed to conventional Type I migration without appealing to more exotic formation scenarios. We therefore note the need to investigate the prevalence of this mechanism in other disc models and for different planet masses. Even if this mechanism does not operate in a wide variety of discs, we suggest that this is just one of a range of mechanisms including disc turbulence \citep{Rein2012} and planetesimal interactions \citep{Chatterjee2015} that may break mean-motion resonances, leading to the observed paucity in \emph{Kepler} systems.

In section \ref{sec:overlib}, we noted that it is possible to break resonances using parametrized migration forces \citep[as in][]{Hands2014}, albeit using separate values of $K_p$ for each planet as opposed to the global $K_p$ that we used previously. Evidently the GS14 mechanism can be reproduced by parametrized disc models, but this did not reduce the high incidence of MMRs in our previous simulations. We believe that this is due to differences in the value of $K_p$. A large fraction of the models that formed compact systems in our previous work had $K_p > 100$, so the eccentricity damping was too strong for the GS14 mechanism to operate. Our results here suggest that a more focused study of the range $10 < K_p < 100 $ is likely to yield a region of the parameter space in which one can both form compact systems and break resonances. This would provide an interesting constraint on the properties of the discs in which compact systems form. We also caution against using constant $K_p$ for all planets in parametrized migration models. This was our previous approach, but Table \ref{tab:timescales} shows that this does not necessarily hold true for varying planet masses. It is unclear how much this affects the operation of the GS14 mechanism, but further studies of how $K_p$ varies with planet mass and how this difference affects resonance breaking may be vital to understanding the formation of compact systems.

An interesting postscript to this discussion is the recent discovery of seven tightly-packed $\sim$Earth-mass planets orbiting the very low mass (0.08M$_{\odot}$) star TRAPPIST-1 \citep{Gillon2017}. In stark contrast to the majority of {\it Kepler} multiple systems, the planets in TRAPPIST-1 appear to be trapped in a long resonant chain \citep{Luger2017}. This points strongly towards migration as the assembly channel for TRAPPIST-1 \citep{Gillon2017}, but also suggests that disc-driven migration operates differently around low mass stars. We make no explicit predictions in this regard, but it is notable that TRAPPIST-1 has higher planet/star mass ratios than considered here (by factors $\sim$2--5). Planet-planet interactions are therefore more dynamically significant, and the threshold eccentricity for overstable librations is also increased (to $e \gtrsim 0.03$--0.04). Moreover, observations of protoplanetary discs around young brown dwarfs and low-mass stars suggest similar disc/star mass ratios to solar-mass stars \citep[e.g.,][]{Scholz2006,Andrews2013,Pascucci2016}, implying that the TRAPPIST-1 planets migrated through a disc with surface density 1--2 orders of magnitude lower than considered here. We therefore speculate that the TRAPPIST-1 system was assembled by migration in the same manner as more massive systems such as Kepler-11, but that stronger planet-planet interactions (relative to the disc-planet torques) inhibit resonance-breaking in planetary systems around very low-mass stars.


\section{Summary}\label{sec:summary}
We have presented high-resolution 2-D hydrodynamic simulations of pairs of planets migrating in the Type I regime. For planet masses 5M$_{\oplus}$ and 10M$_{\oplus}$ convergent migration occurs, and the planets become trapped in the 2:1 mean-motion resonance. Once in resonance the eccentricity of the inner planet grows rapidly, and in all our simulations the planets escape from resonance after a few hundred orbits. The resonances are broken more quickly in highly viscous discs, but the resonance-breaking mechanism operates efficiently even in inviscid discs. The gas disc also damps the planets' eccentricities strongly, and this eccentricity damping leads to overstable librations of the resonant argument. We attribute the resonance-breaking in our simulations to this overstability. 
This mechanism was first proposed by \citet{Goldreich2014}, but has never previously been reported in hydrodynamic calculations. Further simulations are required to understand the range of disc models and planet masses over which this mechanism remains effective. Our simulations also show qualitative changes in the planet-disc interaction with increasing eccentricity, and as a result the resonances are broken rather differently than in analytic calculations. Planets escaping from resonance in this manner can account for the paucity of mean-motion resonances in multi-planet systems observed by \textit{Kepler}, and we conclude that disc-driven migration remains the most plausible means of assembling tightly-packed planetary systems.


\section*{Acknowledgements}
We thank Sijme-Jan Paardekooper, David Velasco, Stephen Fendyke, Chris Nixon, Zo\"e Leinhardt and Walter Dehnen for useful discussions. Additionally, we thank an anonymous referee for comments which greatly improved the manuscript. TOH acknowledges support from a Science \& Technology Facilities Council (STFC) PhD studentship, and from the Swiss National Science Foundation grant number 200020\_162930. TOH and RDA acknowledge support from the Leverhulme Trust through a Philip Leverhulme Prize.  This project has received funding from the European Research Council (ERC) under the European Union's Horizon 2020 research and innovation programme (grant agreement No 681601). This work used the Darwin Data Analytic system at the University of Cambridge, operated by the University of Cambridge High Performance Computing Service on behalf of the STFC DiRAC HPC Facility ({\tt www.dirac.ac.uk}). This equipment was funded by a BIS National E-infrastructure capital grant (ST/K001590/1), STFC capital grants ST/H008861/1 and ST/H00887X/1, and DiRAC Operations grant ST/K00333X/1. DiRAC is part of the National E-Infrastructure.

\appendix
\section{\texttt{FARGO} simulations} \label{sec:fargo}
\begin{figure}
\begin{center}
\includegraphics[width=0.99\linewidth]{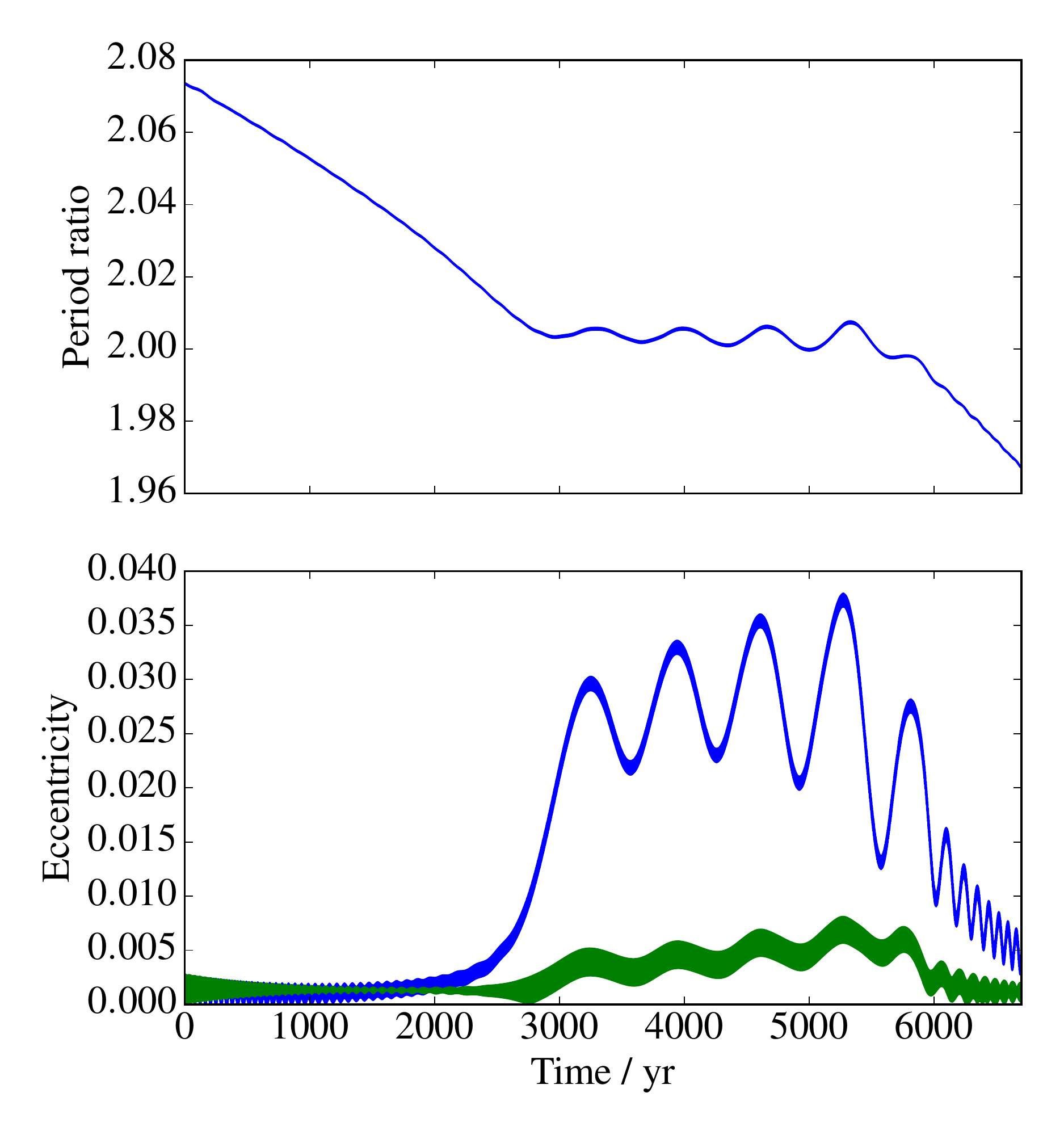}

\caption{As Fig.\ref{fig:1e-3}, but for a simulation using the \texttt{FARGO} code with $\alpha = 10^{-5}$. Compared to the \texttt{PLUTO} simulations, the planets take longer to reach and break through the 2:1 resonance, and the peak eccentricity is slightly higher, but the qualitative behaviour is identical. The quantitative differences are largely due to the differing boundary conditions.\label{fig:fargo}}
  \end{center}
\end{figure}

To verify the numerical accuracy of the \texttt{PLUTO} simulations described in section \ref{sec:setup}, we also performed simulations with a similar setup using the 3D finite-difference code \texttt{FARGO-3D} \citet{Benitez2016}. We used the public version of \texttt{FARGO-3D} in 2D mode, with an identical grid set up and initial conditions to our \texttt{PLUTO} runs. For boundary conditions we used the default \texttt{FARGO} option, which is to extrapolate the surface density and Keplerian velocity profiles across the boundary regions, whilst still using the wave damping scheme of \cite{Borro2006}, whereby any fluctuations in surface density are damped toward their initial value. This is in contrast to our \texttt{PLUTO} runs described in section \ref{sec:bound}, where we used outflow boundary conditions and damped the density to 0 in the wave-damping zones. The \texttt{FARGO} results with $\alpha = 10^{-5}$, are shown in Figure \ref{fig:fargo}.

The \texttt{FARGO} calculations show the same qualitative behaviour as our \texttt{PLUTO} runs: the two planets become trapped in the 2:1 resonance; the eccentricity grows as expected; and then the resonance is broken once a critical eccentricity is reached. Quantitatively, however, all of these processes take longer to occur in the \texttt{FARGO} simulations. The convergent migration is slower and the resonance breaking is then also slower, allowing the eccentricity of the inner planet to reach a marginally higher value. We found these differences to be consistent across all \texttt{FARGO} runs regardless of viscosity.

The origin of this discrepancy in time-scales is primarily the choice of boundary conditions. With our \texttt{PLUTO} outflow boundary conditions the initial surface density profile of the disc is not maintained throughout the simulation, but rather forced to drop smoothly to 0 at the boundaries. This slightly modifies the density gradient at the planets' locations, which in turn leads to small changes in the planets' migration rates (the outer planet migrates more slowly, and the inner planet more quickly, in the \texttt{FARGO} calculations). Test calculations using \texttt{FARGO}-like boundary conditions in \texttt{PLUTO} show very good agreement. There is no ``correct'' answer here: enforcing the initial density profile at both boundaries (the default option in \texttt{FARGO}) prohibits accretion across the inner edge of the grid, and therefore maintains an artificially high disc mass, (particularly for high viscosities); on the other hand, ``outflow'' boundary conditions (as adopted in our \texttt{PLUTO} runs) result in artificial depletion of the disc. However, the two codes show very good agreement when using the same boundary conditions, and the qualitative behaviour of the simulations (escape from resonance due to overstable librations) is unaffected by the choice of boundary condition. 

\bibliographystyle{mnras}
\bibliography{packed}
\appendix

\label{lastpage}

\end{document}